\documentclass[intlimits,twoside,a4paper]{article}

\usepackage{amsmath,amssymb}
\usepackage{graphicx}

\usepackage[T2A]{fontenc}
\usepackage[cp1251]{inputenc}

\usepackage[eqsecnum]{cmpj3}

\issue{2019}{22}{2}{23001}
\doinumber{10.5488/CMP.22.23001}

\title[Dynamics of disease spread. Effect of the characteristic times]%
{Dynamics of disease spread. Effect of the characteristic times%
}

\author[O. Mosbah, N. Zekri, M. Mokhtari, S. Sahraoui]{O. Mosbah\refaddr{label1}, N. Zekri\refaddr{label1}\thanks{Corresponding author, E-mail: nzekri@yahoo.com.},  M. Mokhtari\refaddr{label1,label2}, S. Sahraoui\refaddr{label3} }
\addresses{
\addr{label1} Universit\'e des Sciences et de la Technologie d'Oran Mohamed Boudiaf, USTO-MB, LEPM, BP 1505, \\ El M' Naouar, 31000 Oran, Algeria
\addr{label2} Centre Universitaire de  Tissemsilt, BP 182, Tissemsilt, Algeria
\addr{label3} Le Mans Universit\'e, LAUM, UMR CNRS 6613, Avenue O. Messiaen, 72085 Le Mans cedex 9, France
}

\date{Received November 30, 2018, in final form February 19, 2019}

\begin{document}

\maketitle
\begin{abstract}
Dynamic properties of spreading infection through a heterogeneous population are studied numerically and analytically using a dynamic variant of Watts and Strogatz Small World Network-based stochastic Susceptible-Exposed-Infectious-Removed (SEIR) epidemic model. This model includes the main realistic parameters usually characterizing transmissible diseases, such as the force of infection, latency and infection times. As far as the latency time remains smaller than that of infection, the proportion of infected individuals increases exponentially with time, otherwise an oscillatory behavior appears. This may explain the periodic behaviors in time observed by the health prevention services. It is also shown that periodic epidemiological surveys overestimate or underestimate the dynamics of infection if the survey periods do not exactly correspond to the characteristic times of the infection. Further discussion is provided on the diffusion and relaxation processes involved in this model, and their relation to the infection characteristic time.
\keywords Small World Network, Susceptible-Exposed-Infectious-Removed (SEIR), infection time, latency time, epidemiologic survey, diffusion and relaxation processes
\pacs 05.10.-a, 02.70.-c, 84.37.+q, 51.70.+f, 83.80.Ab, 78.20.-e

\end{abstract}

\section{Introduction}
In developed countries, death rate due to transmissible diseases has fallen sharply over the past century. Already in the late 1960s, it was believed that infectious diseases were being eradicated. Until now, apart from smallpox, transmissible diseases continue to plague despite the advent of vaccination. In addition, in the mid-1970s, new infectious diseases were identified like\textit{ Legionellosis}  \cite{{1},{2}}, \textit{Lyme} disease \cite{{3,3a},{4,4a}}, before other diseases like \textit{Chikungunya} \cite{4,4a}, AIDS \cite{5} or H1N1 Avian flu \cite{6}. These diseases are still not controlled and cause victims. Their control becomes more complex with population mobility and social and environmental changes related to urbanization \cite{1}.

In order to control the behavior of transmissible diseases, mathematical models were proposed for three centuries \cite{{7},{8}}. In the beginning of the last century, Kermack and Mc Kendrick proposed \textit{Susceptible-Infected-Recovered} (SIR) model \cite{9}. In this model,  Susceptible individuals (S) are lacking antibodies or are not/or badly immunized (by vaccine or infection).  Infected individuals (I) are those affected by the virus, whereas Recovered (R) individuals are cured or vaccinated effectively. The parameters S, I and R are usually defined as the proportion of individuals in each class. Later on, more complicated models were proposed including vaccination effects, age structure, etc. \cite{{10},{11}}, and cases where a Recovered individual may become later Susceptible (SIRS). In the case where birth and death rates are neglected, the total number $N$ of population  is assumed constant in time [i.e., $S(t)+I(t)+R(t)=1$]. The SIR model may be described by the following equations:
\begin{equation}
\frac{\rd I}{\rd t}=\beta I S- \gamma I,   
\label{moneq1}
\end{equation}
\begin{equation}
\frac{\rd S}{\rd t}=-\beta I S,
\label{moneq2}
\end{equation}
\begin{equation}
\frac{\rd R}{\rd t}= \gamma I.
\label{moneq3}
\end{equation} 										
Here, $\beta$  is the transmission rate and  $\gamma$  is the recovering rate ($1/\gamma$  is the characteristic recovering period). For a sufficiently large population of Susceptible ($ S \gg \gamma/\beta$), the number of infected individuals~($I$) starts increasing exponentially with time. Stochastic SIR models were proposed to introduce the random character of the transmission and recovering rates due to the population heterogeneity \cite{11}. In order to study the effect of delay on infection, latency time was included in the SIR model [(\ref{moneq1}) to (\ref{moneq3})], the new model is called Susceptible-Exposed-Infectious-Removed (SEIR) model  \cite{11}. This leads to an additional class of individuals ($E$) and equation (\ref{moneq1}) becomes
\begin{equation}
\frac{\rd I}{\rd t}=\sigma E-\gamma I,	
\label{moneq4}
\end{equation}										
where $\sigma$ is the inverse of the latency time $\sigma=1/t_\text{L}$. An additional differential equation is required to describe the evolution of exposed class, and for completes the set    
\begin{equation}
\frac{\rd E}{\rd t}=\beta IS-\sigma E.
\label{moneq5}
\end{equation}
								
\begin{figure}[!t]
\centerline{\includegraphics[width=11 cm]{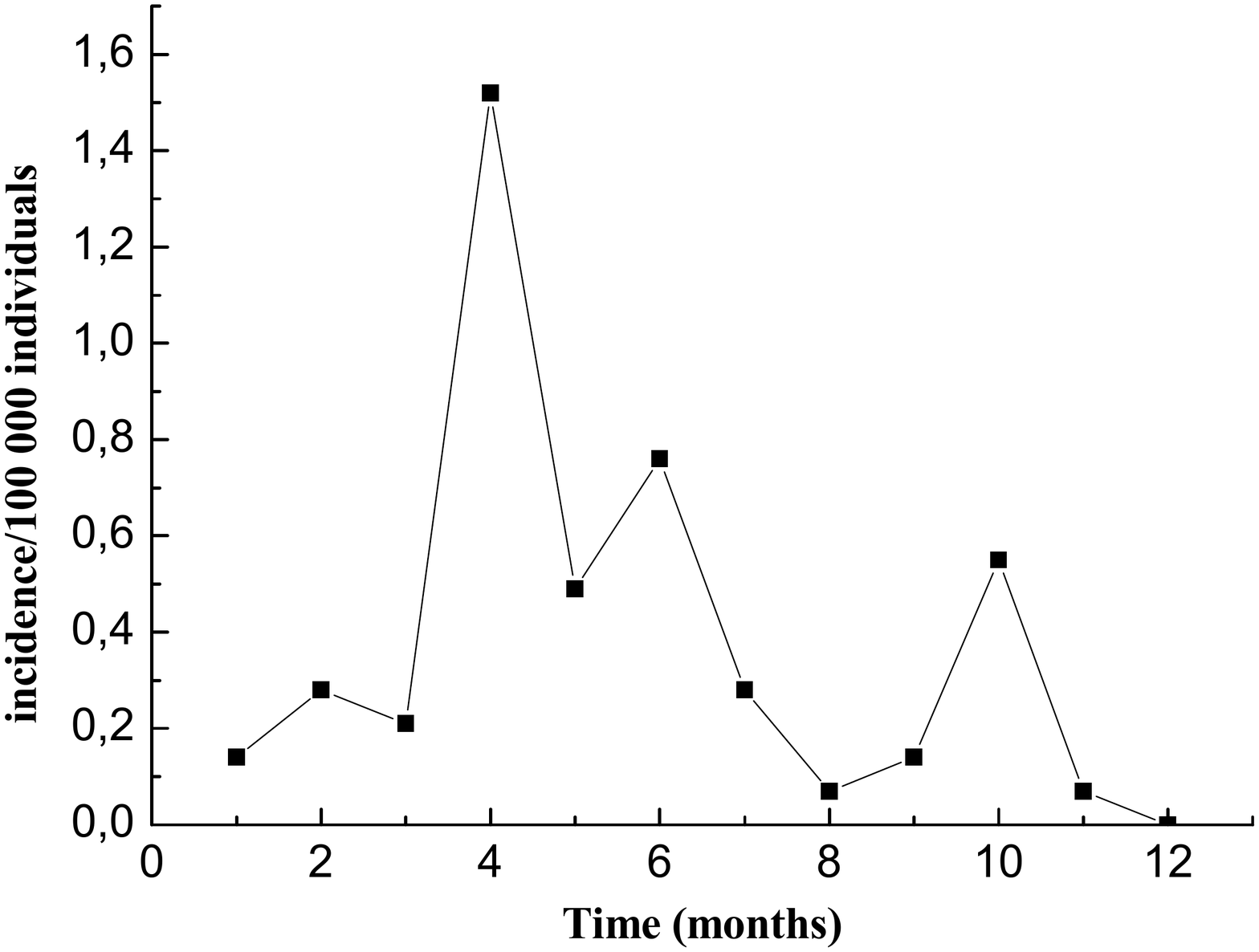}}
\caption{Incidence of measles in the Wilaya of Oran (Algeria) during 2011 (from \cite{28}).}
\label{Fig1:F2H}
\end{figure}

The \textit{Small World Network} (SWN) model proposed in 1998 by Watts and Strogatz \cite{12} reproduces the main features of social networks; i.e., clustering and a logarithmic size-dependent mean free path. \textit{Small World Network} and \textit{Scale Free} models have been used extensively to study human disease and other spreads (computer virus, rumor, etc.) \cite{13,14,15,16,17,18,19,20,21,22}. Zekri and Clerc found that infection grows exponentially \cite{23} above the percolation threshold \cite{24}. They used a heterogeneous SIR model based on a variant of Watts and Strogatz Networks. The rate of the exponential growth was found to increase logarithmically with $p$, $p$ being the initial proportion of Susceptible [$p=S(0)$]. However, this model \cite{23} neglected some realistic parameters like the characteristic times of disease (latency and infection times), and the force of infection. Further models including various effects such as vaccination and strategies for disease eradication \cite{13, 16}, connection length \cite{17,18,19,20,21,22,25}, fluctuations and correlations \cite{22}, and cost control of epidemics \cite{13, 17} have been proposed. Recently a distribution of Latency time was used to study the global stability of spread dynamics \cite{26}. Latency time can significantly modify the disease spread dynamics. In particular, a small latency time appears as a slight delay of the spread, but larger values might lead to disease extinction. In \cite{20}, these characteristic times were included in the model, but their effect on spread dynamics was not examined. Another work introduced the infection time to provide a dynamic character to the disease behavior, and to study its effect on the percolation threshold \cite{25}. However, the effect of this time on disease spread was not investigated. More recently, these characteristic times were introduced in order to simulate historical data on influenza \cite{27} (here, the data of survey was collected every week). The time evolution of the infection has shown a nonlinear increase with plateaus of two weeks. Another survey investigation concerning measles disease was realized during 2011 with periods of two weeks. The infection behavior was even found oscillatory \cite{28} (see figure \ref{Fig1:F2H}). The exponential increase was not observed in the later investigation due to the large period of survey. Therefore, simulations fail in reproducing the observed spread dynamics if the period of survey is arbitrarily chosen.

In the present work, an SEIR model based on a dynamic variant of SWN network is used to study:  a)~the effect of latency and infection times on the disease spread dynamics, and  b) the comparison between the continuous (daily) recording and larger surveys recording periods. Indeed, the health monitoring services record the infection during given periods, which could influence the data. 

\section{Description of the model}

The original version of the \textit{SWN}  model proposed by Watts and Strogatz was based on an L-sized regular $1\text{D}$ network with periodic boundary conditions ($2\text{D}$ regular network can be re-indexed into $1\text{D}$ network). Each node of the network is connected to $K$ nearest neighbors, and each bond is rewired, with probability $\varphi$, to a new node randomly and uniformly chosen from the network \cite{12}. A number of problems have been found with this model, the most serious among them is that of calculating the average distance between vertices (or the mean path length) in the limit of vanishing probabilities ($\varphi=0$). In order to avoid these problems, Newman and Watts proposed later a variant of this network model \cite{29}, where bonds are not rewired but additional long-range connections (called short-cuts) beyond the nearest neighbors are added randomly through the network with a probability $\varphi$. For very small probability, the network remains regular. It becomes irregular as the probability ($\varphi$) increases. In the model of Watts and Strogatz \cite{12}, the rewired short-cuts made the network composed of local clusters, whereas that of Newman and Watts \cite{19} is a single local cluster.

Newman and Watts examined the percolation threshold $p_\text{c}$ [i.e., the minimum initial proportion of susceptible individuals $S(0)$ allowing the largest connected cluster to cover a significant fraction of the entire network, where an epidemic ensues \cite{23}].  The threshold ($p_\text{c}$) was found to rise but did not vanish as the probability ($\varphi$) of short-cuts approaches the unity [$p_\text{c} (\varphi=1)\neq 0$], as shown in figure~7 of \cite{29}.  This curve seems analytic, and thus percolation threshold is expected to continue decreasing for $\varphi>1$. Therefore, the parameter $\varphi$ can be regarded as the average number of short-cuts per node (individual) instead of their probability. The curve (figure~7 of \cite{29}) has a limit (maximum value $\varphi_\text{max}$) at which the percolation threshold vanishes. In this limit, the network model becomes a free scale-like (the presence of any infection leads to the epidemic phase). In real human networks, each individual can meet much more than one individual (a sub-group) to whom the disease can be transmitted. Processes with a number of long-range contacts per individual larger than one ($\varphi>1$) have been already investigated for measuring concurrency in systems characterizing human contacts \cite{30}. In the case of free scale networks, the sub-group covers a large proportion of the network \cite{31, 32}. These networks are very heterogeneous and are composed of a few highly connected nodes (also called hubs or hot points). They characterize most of social networks such as the World Wide Web and computer viruses. In the case of computer viruses, the hot points are actually Google, Yahoo, etc.

The present SEIR epidemic model is based on the variant of SWN model described above, which is a generalization of the epidemic models from the initial Watts and Strogatz SWN to the free scale networks. The node represents an individual, which is initially ($t=0$) either in the class of Susceptible with a probability $p=S(0)$ or recovered (or immunized) with a probability  $1-p=1-S(0)$. For $p\neq 0$, the system becomes composed of clusters of Susceptible nodes of different sizes connected together \cite{34}. If the system size is infinite, the size of the largest cluster diverges at the percolation threshold \cite{24}. The parameters $K$ and $\varphi$ are chosen to model childhood networks. Nearest neighbor connections correspond to contacts of an individual within his or her family or neighborhood, whereas short-cut connections represent contacts with individuals beyond its neighbors in a public place (like a school). These later contacts are different from those of nearest neighbors, since the virus is actually transmitted through the school instead of at home.  The school is then a ``virtual'' node through which long-range connections between real nodes (individuals) occur. The network has a spatial organization, since the indices of the nodes are incremented from a site to its nearest neighbor, and a kernel defines here the distances between nodes.

\begin{figure}[!b]
\centerline{\includegraphics[width=12 cm]{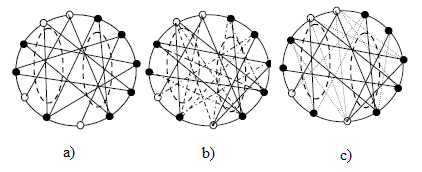}}
\caption{1D SWN of 12 nodes with  $K=2$ and $\varphi = 2$, and $p=2/3$.   a)  	Short-cuts are frozen in time, b)~same as a) after $\delta t$ with dashed lines short-cuts  c)  The same as a)  after $2 \delta t$ with dotted lines short-cuts. a) (solid lines) are shown in b) and c) for comparison. Dashed ellipses are ``virtual'' nodes.}
\label{Fig2:F2H}
\end{figure}

The present network model is shown in figure~\ref{Fig2:F2H}, where short-cuts are links beyond neighbors. The Full circles correspond to Susceptible nodes, and empty circles correspond to Recovered (or immunized) ones. Unlike the network used by Zekri and Clerc \cite{23}, where all connections are frozen in time [figure~\ref{Fig2:F2H}~(a)], in the dynamic network considered here, the short-cut connections change to figure~\ref{Fig2:F2H}~(b) after a step $\delta t$ (taken here as one day) and to figure~\ref{Fig2:F2H}~(c) after $2 \delta t$. In each time step, each node has in average $\varphi$ new short-cut connections to other nodes randomly and uniformly chosen from the network. In a public place (``virtual'' node), an individual does not necessarily meet the same individuals every day. The neighbor connections are rather fixed in time.

In real life, the number of connections fluctuates from a site to another one depending on the size of each family, neighborhood or the public place of the encounters as well as the age of the population under consideration. These fluctuations can be generated using a Gaussian probability distribution of connections, assuming the number of short-cut connections per site uncorrelated \cite{33}. A Gaussian distribution of the number of short cuts was found, in a previous work to shift the percolation threshold towards higher thresholds \cite{34}. It is found here (not shown) that the fluctuations of the number of short-cuts shift the growth rate towards lower values. For simplicity, the number of connections are chosen typically to $K = 2$ (an average number of contacts between neighbor individuals of similar ages), and  $\varphi=40$ (i.e., the average number of kids that an individual can meet in a school class). The connectivity is then $z=K+ \varphi =42$. Note that the connectivity does not affect the disease general behavior. It modifies the rate of growth and the percolation threshold.

Let us describe the spread process (simulation algorithm) through the present SEIR model. Initially ($t=0$) anode chosen randomly inside the largest cluster of Susceptible nodes changes its status to the Infectious class, and starts the infection process. Time is incremented by steps  $\delta t$ for all the nodes of the network, where Susceptible nodes connected to an Infectious one change their status to the class of Exposed with a probability $q$  (the force of infection). The virus invades only the body of each Exposed node during the latency period $t_\text{L}=1/\sigma$, where it cannot contaminate its connected Susceptible nodes. Each Exposed node has its own clock, which ends after the period $t_\text{L}$. After this period, the status of the Exposed node changes to the Infectious class. Each Infectious node has its own clock ending after the period $t_\text{I}$, then its status changes to the Recovered class. The contamination process continues until all the nodes of the cluster change to the Recovered class. The nodes in the Recovered class do not participate in the infection process. The update process after each step $\delta t$  is realized sequencially as in cellular automaton.
The total period of infection defines the recovery rate $\gamma$ introduced in (\ref{moneq3}) ($t_\text{L}+t_\text{I}=1/\gamma$).  In the previous work \cite{23}, the latency time was neglected ($t_\text{L}=0$), the infection period was taken as the unit time ($t_\text{I}=\delta t=1$ day), and the force of infection was maximum ($q=1$). The characteristic times ($t_\text{L}$ and $t_\text{I}$) depend on the disease, age and health of the individual \cite{35}. For influenza, the latency time varies generally from $1$ to $7$ days (the incubation period is usually $2$ days). The infection time is generally $7$ days for an adult and $21$ days for children less than  $12$ years old.  For measles, the latency time tends asymptotically to an average between $10$ to $12$ days, and can be extended to $18$ days from the exposition to the appearance of the fever. The infection time is at maximum of $4$ days extending to $6$ days. In this work, the force of infection, infection and latency time are chosen arbitrarily to study their influence on the spread dynamics.

\section{Results}

Simulations are realized on sample networks of $100$ $000$ nodes (representing individuals). Infected individuals are initially ($t=0$) either Latent (their proportion is $I_\text{L}$) or Infectious (their proportion being~$I_\text{I}$). The total proportion of infected individuals is then $I=I_\text{L}+I_\text{I}$. In order to minimize fluctuations, data are averaged over $1 000$ configurations. In what follows, the force of infection and the initial proportion of Susceptible individuals are respectively fixed to $q=p=0.3$ otherwise specified (these parameters are chosen to ensure the spread, i.e., $p>p_\text{c}$). The results can be presented either continuously (daily) or using surveys (the data are accumulated) during periods $\theta > \delta t$. The effect of the infection parameters on the spread dynamics is examined using the continuous (daily) data. The periodic surveys are then compared to the continuous data in order to show the effect of  $\theta$  on the deviation from spread dynamics.

\subsection{Influence of the infection characteristic times on disease spread dynamics}

Figure~\ref{Fig3:F2H}~(a) shows the evolution with time of the proportions of Infected ($I$), Latencies ($I_\text{L}$) and Infectious ($I_\text{I}$) individuals for $t_\text{I}=3$ days and $t_\text{L}=2$  days.  The exponential behavior predicted by equation~(\ref{moneq1}) is confirmed for both types of Infected individuals, with a delay of $t_\text{L}$ for Infectious ones $I_\text{I}$ compared to the Latencies $I_\text{L}$:
\begin{equation}
I_\text{I,L} \propto \text{e}^{(t/\tau)}.	
\label{moneq31}
\end{equation}

The same growth exponent is observed (within the statistical errors) for both kinds of infected individuals. This exponent corresponds to the growth characteristic time $\tau$.

Let us examine the physical meaning of the growth time $\tau$.  In figure~\ref{Fig3:F2H}~(a), a maximum proportion of infected individuals $I_\text{max}$ is reached. Beyond this maximum, a decrease appears due to the lack of Susceptible individuals, and the finite size of the sample. This maximum proportion is independent of the size (the size effect is not shown here). For an infinite size, this maximum is reached asymptotically  ($t\rightarrow\infty$).  This behavior appears similar to the relaxation process. Consequently, the proportion of Susceptible individuals decreases exponentially with time during the virus invasion. Here, the sample of Susceptible individuals can be considered analogous to a condenser which is discharged with time through a resistance. The charge corresponds to the proportion of Susceptible individuals, and the resistance to the diffusion of the virus through the population sample. The proportion of Susceptible individuals is then expected to decrease as:
\begin{equation}
S(t)=I_\text{max}-I(t)=I_\text{max} \re^{-t/t_\text{r} }.	
\label{moneq32}
\end{equation}

Here, $t_\text{r}$  is the relaxation time of this discharge. Metzler et al. \cite{{36},{37}}  have related the relaxation process to the diffusion phenomenon. Anomalous relaxation was observed in dielectrics \cite{38} and viscoelasticity~\cite{39},  and was attributed to the sub-diffusion phenomenon \cite{{36},{37}}. In case of sub-diffusion, equation (\ref{moneq32}) becomes:

\begin{figure}[!t]
\centerline{\includegraphics[width=14 cm]{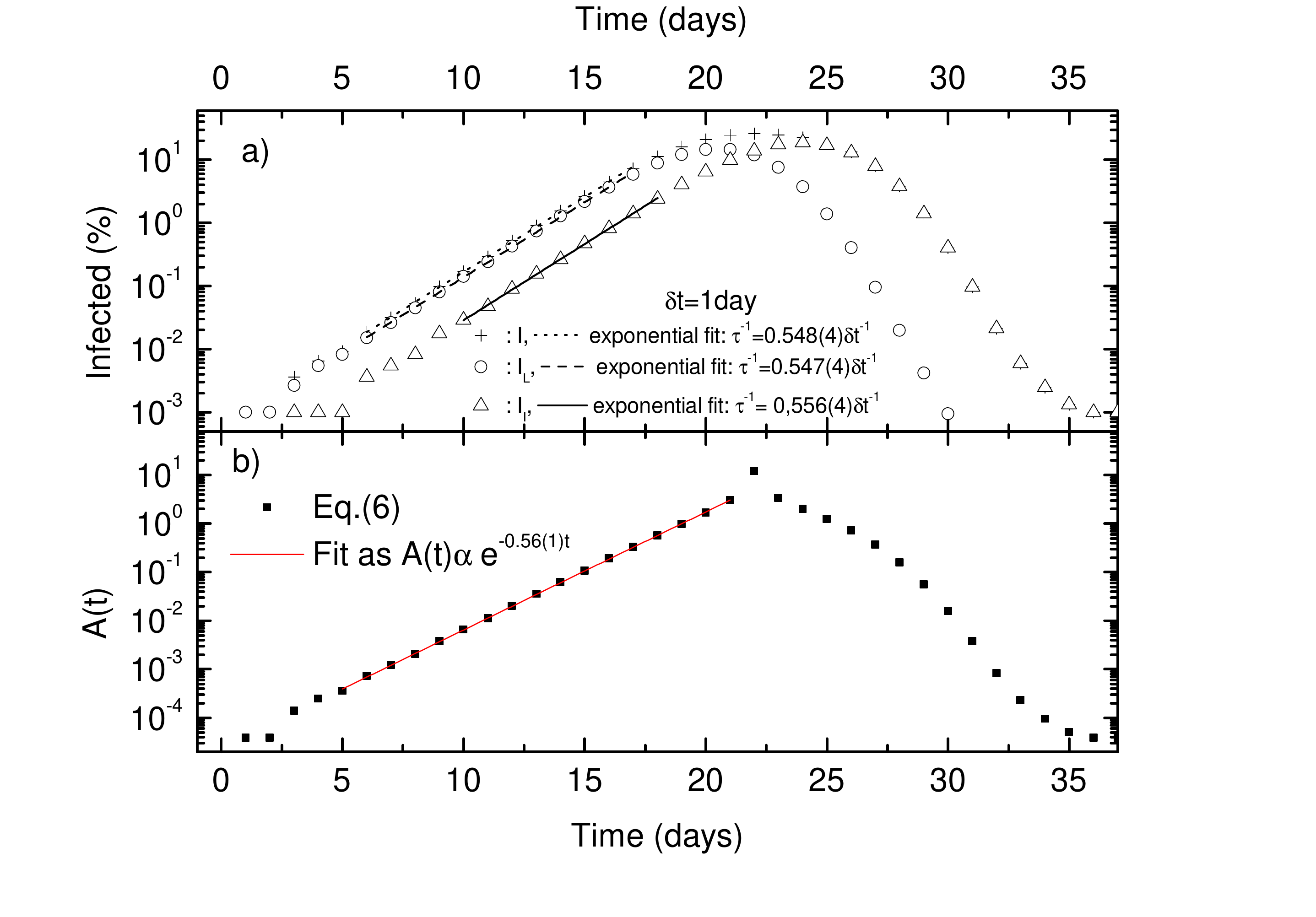}}
\caption{Semi-logarithmic plots of a) time evolution of infected $I$, latencies $I_\text{L}$ and infectious individuals~$I_\text{I}$, and b) $A(t)$. $ t_\text{I}=3$ days, and $t_\text{L}=2$  days.}
\label{Fig3:F2H}
\end{figure}
							
\begin{equation}
S(t)=I_\text{max}-I(t)=I_\text{max} \re^{-(t/t_\text{r} )^\beta};\,\,\,\,\,  \beta<1.	
\label{moneq33}
\end{equation}

On the other hand, it has been found that the disease spread in a SWN network is an exponential super-diffusion \cite{23}.  A new kind of relaxation is thus expected for the present spread. In order to determine the kind of relaxation, equation (\ref{moneq32}) can be re-written as follows:
\begin{equation}
A(t)=\ln{ \frac{I_\text{max}}{I_\text{max}-I(t)}}=t/t_\text{r}\,.		
\label{moneq34}
\end{equation}

The quantity $A(t)$ behaves thus linearly with time for a normal relaxation, and as a power-law in the case of anomalous relaxation due to the sub-diffusion (\ref{moneq33}).  It is then interesting to examine its behavior in the present SWN model.  In figure~\ref{Fig3:F2H}~(b),  the time dependence of the quantity $A(t)$ is presented in a semi-logarithmic plot.  An excellent linear fit with $ R^2>0.999$ is observed,  indicating, in contrast to equations (\ref{moneq32}) and (\ref{moneq33}),  the following kind of relaxation 
\begin{equation}
I_\text{max}-I(t) \propto \re^{\re^{-t/t_\text{r}}}.	
\label{moneq35}
\end{equation}

By analogy with the super-diffusion process of disease spread,  we call this kind of relaxation \textit{super relaxation}. Within the statistical errors, the relaxation time ($1/t_\text{r}=0.56\pm 0.01$  day$^{-1}$) obtained from the linear fit [equation~(\ref{moneq35})] in figure~\ref{Fig3:F2H}~(b) seems to be compatible with the characteristic growth time $\tau$ ($1/\tau =0.556\pm 0.004$  day$^{-1}$) as shown in figure~\ref{Fig3:F2H}~(a),  which means that  the growth time of (\ref{moneq31}) is in fact a relaxation time. 

Let us examine the growth exponent ($1/\tau$) dependence on the disease parameters. The exponent was previously found to be proportional to $\log(p)$ \cite{23}. This exponent may be influenced in the present model by five parameters  ($p$, $q$, $z$, $t_\text{L}$ and $t_\text{I}$).

In the absence of latency ($t_\text{L}=0$), and for $t_\text{I}= \delta t$ \cite{23},  all infected individuals are systematically in their infectious phase. For a given connectivity $z$, the proportion of infected individuals varies with time as ($t=n_t \times \delta t$): 
\begin{equation}
I(n_t )=I(n_t-1) \times q z p=I_0 (q z p)^{n_t }.	
\label{moneq36}
\end{equation}

$I_0$ being the initial number of infectious individuals.  Equation (\ref{moneq36}) yields an exponential growth in the form of (\ref{moneq31}) with an exponent:
\begin{equation}
1/\tau	= \ln{(q z p)}.
\label{moneq37}
\end{equation}
											
This explains the logarithmic behavior of the growth exponent observed in \cite{23}.  For $t_\text{I} > \delta t$, the new and old infectious individuals are accumulated in the infectious phase during $t_\text{I}$, and (\ref{moneq36}) becomes:
\begin{equation}
\begin{gathered}
I(n_t ) = I(n_t-1)q \overline{z} p+ \sum_{j=1}^{\frac{t_\text{I}}{\delta i}-1} I(n_t-j)-I(n_t-t_\text{I}/\delta t) \\
            =I(n_t-1)(1+q z p)+\sum_{j=2}^{\frac{t_\text{I}}{\delta i}-1} I(n_t-j) -I(n_t-t_\text{I}/\delta t).
\end{gathered}	
\label{moneq38}
\end{equation}
	
Equation (\ref{moneq38}) can be rewritten as:
\begin{equation}
I(n_t )=I_0 (qzp+1)^{n_t } \big\{1+(n_t-1) (qzp+1)^{-(t_\text{I}/\delta t-1)}+O\big[(qzp+1)^{-2(t_\text{I}/\delta t-1)} \big] \big\}.
\label{moneq39}
\end{equation}

For $t_\text{I}\gg \delta t$, the first term is dominant in (\ref{moneq39}) and the growth exponent is 
\begin{equation}
1/\tau \approx \ln(q z p+1).	
\label{moneq310}
\end{equation}
										
Therefore, for large infection time, the growth process remains exponential and its exponent is logarithmically dependent on $q z p$. It was assumed here that the proportion of Susceptible $S(t)$ remains constant during the whole period of spread. However, this proportion decreases during the growth (\ref{moneq35}). Actually, for $t_\text{I}= \delta t$, $q z p$ new Susceptible become infected at each time step. If the initial proportion of Susceptible is $p=S(0)$, the proportion of Susceptible behaves with time as ($\alpha= {qzp}/{N}$):
\begin{equation}
p(n_t )=  \begin{cases}
p_0  \sum_{j=0}^{n_t} (-\alpha)^j  \frac{(n_t !)}{(n_t-j!)}  & \text{ for $t_\text{I}= \delta t$},\\ 
\\
p_0 \sum_{j=0}^{n_t}(-\alpha+1/N)^j   \frac{(n_t !)}{(n_t-j!)}   & \text{for $t_\text{I} > \delta t$}.
\end{cases}	
\label{moneq311}
\end{equation}
				
\begin{figure}[!t]
	\vspace{-6mm}
\centerline{\includegraphics[width=14.3 cm]{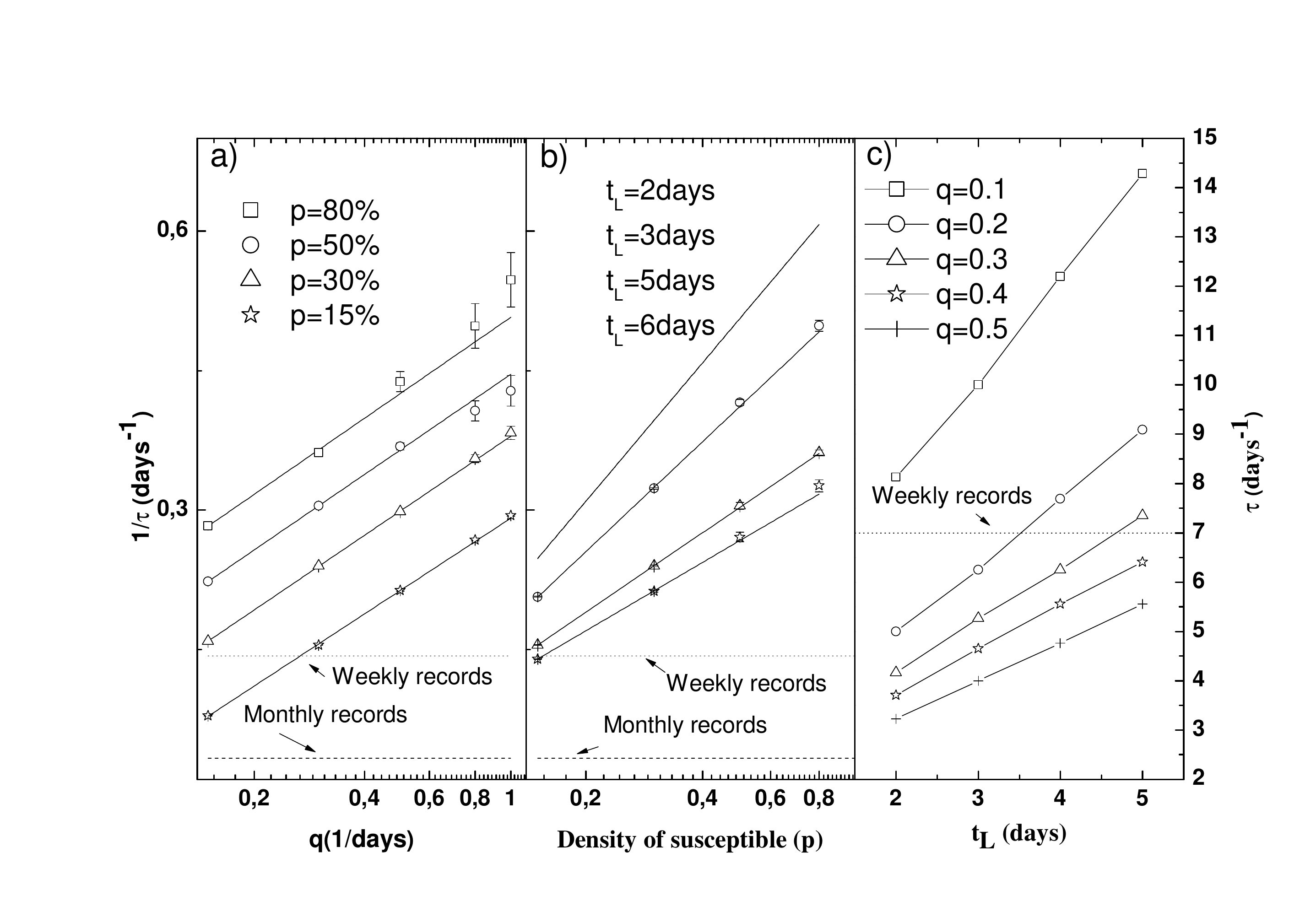}}
\caption{Semi-logarithmic plots of $1/\tau$  vs.: a) $q$ for different $p$ ($t_\text{L}=5$  days), b) $p$ for various $t_\text{L}$ ($q=0.3$), and c) $t_\text{L}$ for various $q$ ($p=0.15$).  Data are compared with weekly and monthly surveys. $t_\text{I}=7$ days. Continuous lines are logarithmic fits of data.}
\label{Fig4:F2H}
\end{figure}

In figures~\ref{Fig4:F2H}~(a) and (b), the growth exponent $1/\tau$  seems to depend logarithmically on the strengths of $p$ and $q$ confirming the behavior of equations [(\ref{moneq37}) and (\ref{moneq310})] with an excellent fit of data for both figures  ($R^2 >0.99$). Since the proportion of Susceptible varies with time [(\ref{moneq35}), (\ref{moneq311})], the growth exponent depends on an effective proportion $p_\text{eff}$ instead of the initial one ($p$). 

\begin{figure}[!b]
	   \vspace{-8mm}
	\centerline{\includegraphics[width=14.3 cm]{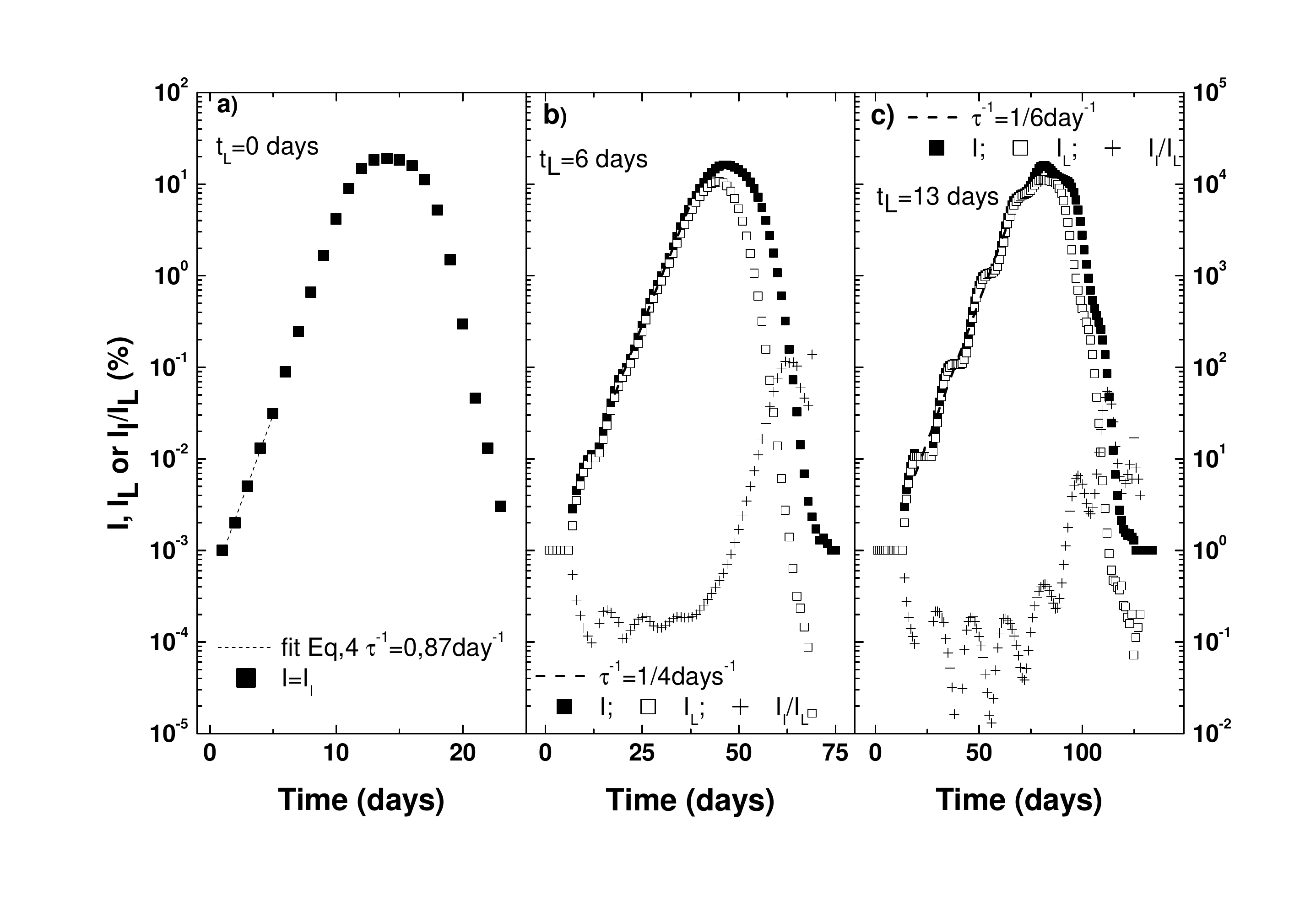}}
	\caption{Semi-logarithmic plots of: $I$ (full squares), $I_\text{L}$ (open squares) and $I_\text{I}/I_\text{L} $ (+) vs. time for different latency periods. a) $t_\text{L}=0$  day,  b) $t_\text{L}=6$  days,  c) $t_\text{L}=13$  days. Infection time is $6$  days in both figures.}
	\label{Fig5:F2H}
\end{figure}

	Now, let us introduce the latency time $t_\text{L}$  on the spread. The characteristic time $\tau$  of the growth appears to increase linearly with the latent period $t_\text{L}$ [figure~\ref{Fig4:F2H}~(c)],  as if this period were added to the characteristic time.  As expected,  the latent period shifts the infection, and slows down its growth dynamics.  Unlike the slowing effect caused by the latency time $t_\text{L}$, the infection time $t_\text{I}$ strengthens the spread and broadens the infection duration. One expects then a competition between these two effects.  Indeed,  in the limit of very large latency times ($t_\text{L} \gg t_\text{I}$), the infectious individuals will obviously disappear,  before appearing during the infection period.  This may induce periodic plateaus and an oscillatory behavior in the evolution of the proportion of \textit{Latencies} and \textit{Infectious} respectively.  The plateaus and oscillatory behaviors are clearly observed in figures~\ref{Fig5:F2H}~(b) and (c),  for $t_\text{L}\geqslant t_\text{I}$, compared to figure~\ref{Fig5:F2H}~(a), where the infected individuals are systematically infectious ($t_\text{L}=0$) and $I=I_\text{I}$. The infection period introduces then a memory effect, since the number of newly infected individuals at a time $t$ is obtained not only from the new infectious ones at time $t- \delta t$, but also from those appearing at different times up to $t-t_\text{I}$.  After the period $t_\text{L}$, the latencies become infectious and the newly infected individuals at time $t-t_\text{L}$ leave the latency class at time $t+\delta t$. The evolution of the infection is thus governed by the following differential equations for latencies and infectious
\begin{equation}
\frac{\partial I_\text{L} (t+\delta t)}{\partial t}=[I_\text{I} (t)- I_\text{I} (t-t_\text{L} ) ] q z p,	
	\label{moneq312}
\end{equation}
\begin{equation}
\frac{\partial I_\text{L} (t-t_\text{L})}{\partial t}=\frac{\partial I_\text{I} (t+\delta t)}{\partial t}=[I_\text{I} (t-t_\text{L} )-I_\text{I} (t-t_\text{L}-t_\text{I} ) ].
\label{moneq313}
\end{equation}
					
These relations show the delay $t_\text{L}$ between Latencies and Infectious individuals. If we restrict ourselves to the latencies, (\ref{moneq312}) can be re-written as follows:
\begin{equation}
\frac{1}{I_\text{L}} \frac{\partial I_\text{L} (t)}{\partial t}=\frac{I_\text{I} (t)}{I_\text{L} (t)} \bigg[1- \frac{I_\text{I} (t-t_\text{L} )}{I_\text{I} (t) }\bigg]qzp=\frac{I_\text{I} (t)}{I_\text{L} (t) } \bigg[1-\frac{I_\text{I} (t-t_\text{L} )}{I_\text{L} (t-t_\text{L} )}\bigg] q z p.
\label{moneq314}
\end{equation}
		              
If the ratio $I_\text{I} (t)/I_\text{L} (t)$ (which depends on $q z p$) is constant in time, the exponent $1/\tau$  becomes
\begin{equation}
\frac{1}{\tau}=\frac{I_\text{I} (t)}{I_\text{L} (t)} \bigg[1- \frac{I_\text{I} (t-t_\text{L} )}{I_\text{L} (t-t_\text{L} ) }\bigg] q z p.	
	\label{moneq315}
\end{equation}
						               
The fitted value of the growth exponent $1/\tau$ is $0.25\pm 0.01$  day$^{-1}$ in good agreement with the analytic value   $ \big[{I_\text{I}}/{I_\text{L} (1-\frac{I_\text{I}}{I_\text{L}} )}\big] q z p$  in (\ref{moneq315}) with $I_\text{I}/I_\text{L} \approx 0.18$.

  The positivity of the solution is always fulfilled since in all spreading cases, the number of latency individuals must be greater than that of infectious individuals.


\subsection{Effect of the survey periods on the evolution behavior}

The health monitoring services carry out periodic surveys (Daily, weekly, monthly\dots) for the estimation of the infection dynamics. This estimation is influenced by the survey period $\theta$.  It is important to know to what extent the survey data corresponds to the infection dynamics. 

 If the survey daily reports all newly infected individuals  during a period $\theta$, the estimated proportion of infected individuals is:
\begin{equation}
I_\text{S} (t_0;  n\theta)=\int_{t_0+(n-1)\theta}^{t_0+n\theta} \frac{\partial I_\text{L}(t)}{\partial t} \rd t    \quad   \textrm{with}   \quad  n>1.   	
\label{moneq316}
\end{equation}
		
The survey data do not seem to depend on the initial time $t_0$. In figure~\ref{Fig6:F2H}, the evolution with time of the proportion of infected individuals ($I_\text{I}$, $I_\text{L}$ and $I_\text{L}+I_\text{I}$) is compared to the estimated survey data  $I_\text{S}$ for different periods $\theta$. Two different cases are presented: $t_\text{L}>t_\text{I}$  [figure~\ref{Fig6:F2H}~(a)], where the plateaus and oscillations appear, and $t_\text{L}<t_\text{I}$  [figure~\ref{Fig6:F2H}~(b)]. In all cases, the estimated number $I_\text{S}$ increases as the period $\theta$  is increased. When $\theta$ varies (compared to $t_\text{L}$  and $t_\text{I}+t_\text{L}$), the survey data $I_\text{S}$  behave as follows:
 
\begin{figure}[!t]
\centerline{\includegraphics[width=14 cm]{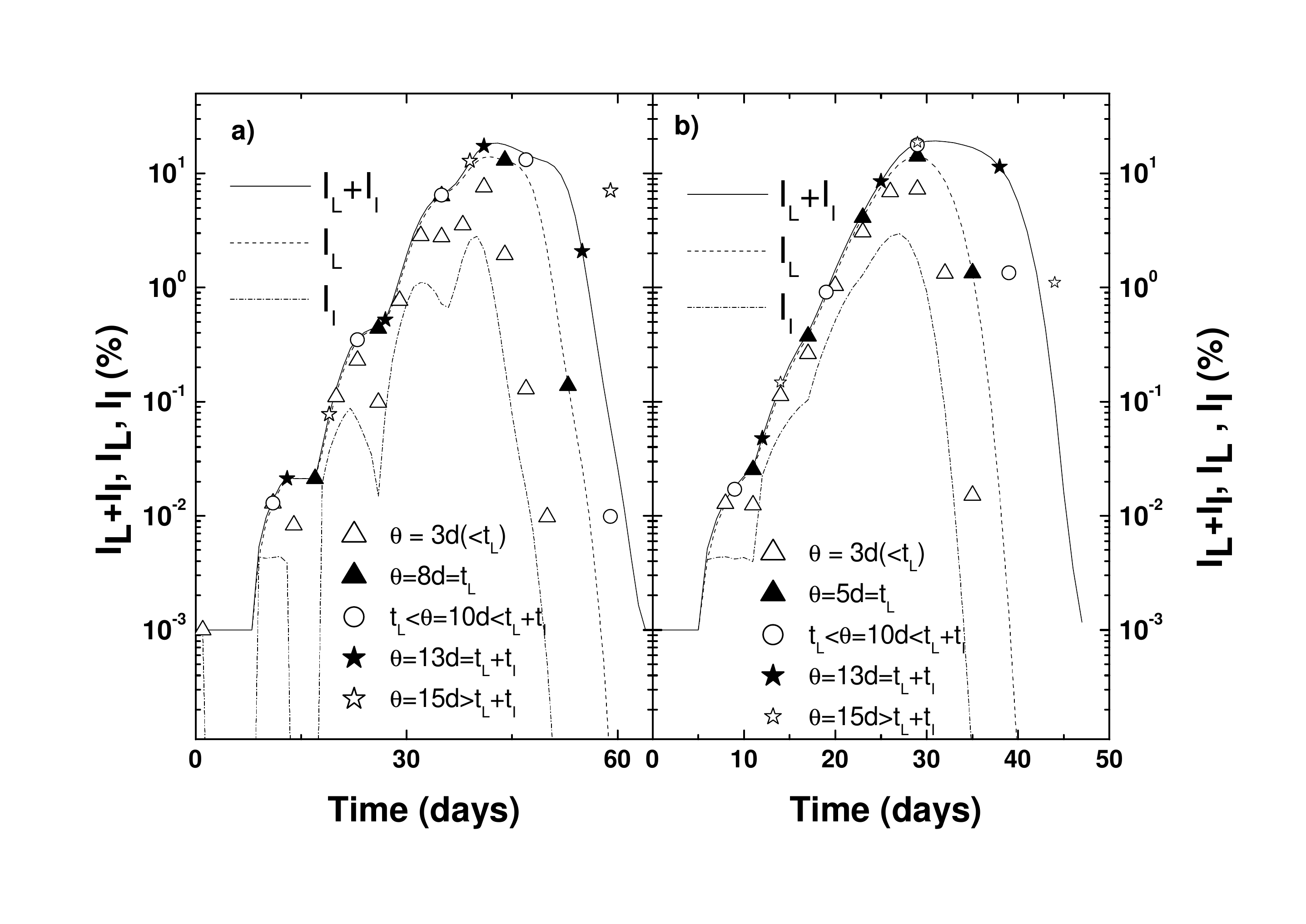}}
\caption{Semi-logarithmic plot of temporal evolution of $I$, $I_\text{L}$   and $I_\text{I}$  (curves) and  $I_\text{S}$ (symbols)  for $p= 0.2$, and $q=0.5$ and different survey periods $\theta$, a) $t_\text{L}=8$  days and $t_\text{I}=5$  days, b) $t_\text{L}=5$  days, $t_\text{I}=8$~days.}
\label{Fig6:F2H}
\end{figure}
\begin{itemize}

              \item For $\theta<t_\text{L}$,  the survey data are underestimated  $I_\text{S}<I_\text{L}$,  since not all Latencies are recorded during the period $\theta$. In the limit $\theta \rightarrow 0$, $I_\text{S}$ coincides with the proportion of new infected individuals, and behaves proportionally to $I_\text{I}$  from (\ref{moneq316}).

	\item For $ \theta=t_\text{L}$  (filled triangles), $I_\text{S}=I_\text{L}$. The reported data correspond exactly to the Latencies.

	\item For $t_\text{L}<\theta<t_\text{L}+t_\text{I}$, $I_\text{I}+I_\text{L}>I_\text{S}>I_\text{L}$. In this case, the survey accounts for all individuals in their latency period, and also part of those in their infectious period.

	\item For $\theta=t_\text{L}+t_\text{I}$, $I_\text{S}=I_\text{L}+I_\text{I}$. The reported data correspond exactly to the total proportion of infected individuals (filled stars).

	\item For  $\theta>t_\text{L}+t_\text{I}$, $I_\text{S}>I_\text{L}+I_\text{I}$.  The survey overestimates the infection since recovered individuals are accounted with the infected ones.

\end{itemize}

Therefore, the monitoring services must choose the survey period either at the  $\theta=t_\text{L}$  or $\theta=t_\text{L}+t_\text{I}$, otherwise $I_\text{S}$ deviates from the real infection dynamics. The aim of this part of the study is to warn the operational of health on the error made if a survey with an arbitrary period is realized. We do not intend to estimate practically the right infection dynamics from the surveys, although a daily survey yields the right disease dynamics. There are different biological and epidemiological methods to estimate the latency and infection times. These times are actually around $7$ days for measles. Then, weekly surveys are closer to the right infection dynamics for this disease. Monthly surveys overestimate this dynamics. A method can be proposed to avoid the over/under estimation of the infection dynamics. We propose to start with a daily survey during the two first weeks, then to compare the data for different periods in order to find the one closer from daily survey. In the results of figure~\ref{Fig6:F2H}, the survey assumes that each newly infected individual is recorded systematically at time by the health monitoring services (i.e., recording is assumed with no delay). In case of delay fluctuations in recording, the survey data become completely different from the real infection dynamics. 

\vspace{-4mm}
\section{Conclusion}
\vspace{-2mm}
In this paper, the dynamics of disease spread was examined numerically using the SEIR model within the framework of a dynamical SWN-like Network. The growth exponent $1/\tau$ was found to correspond to a super-relaxation time for this kind of lattices. Its logarithmic dependence on the proportion of Susceptible~$p$, observed previously \cite{23} is also confirmed for the force of infection $q$. The latency time~$t_\text{L}$ slows down the disease spread, and the characteristic time $\tau$ increases linearly with $t_\text{L}$.

When the latency time becomes larger than the infection one ($t_\text{L}>t_\text{I}$), the dynamics of spread shows periodic behaviors with plateaus of latent individuals and oscillations of the infectious ones.

Finally, the estimation data of periodic surveys carried out by the health monitoring services are compared to the infection dynamics. The survey data represent the real spread dynamics only if the period $\theta$ coincides with the latency one $t_\text{L}$ or with the total infection period ($t_\text{L}+t_\text{I}$). The periodic survey should then account for disease characteristic times.

\vspace{-4mm}
\section*{Acknowledgements}
\vspace{-2mm}
Simulations were realized using IBN BAJA HPCat USTO-MB.

\newpage
\ukrainianpart

\title{Динаміка розповсюдження хвороб. Вплив характерних періодів часу}
\author{O. Мосбах\refaddr{label1}, Н. Зекрі\refaddr{label1},  M. Мохтарі\refaddr{label1,label2}, С. Сахраві\refaddr{label3} }
\addresses{
	\addr{label1} 
	Унiверситет науки i технологiй iм. Мохамеда Будiафам, Оран, USTO-MB, LEPM, BP1505, 31000 Оран, Алжир	
	\addr{label2}Унiверситетський центр Тiссемсiлту, BP182, Тiссемсiлт, Алжир
	\addr{label3} 
	Університет Ле-Ман, LAUM, UMR CNRS 6613,  72085 Ле Ман, Франція}

\makeukrtitle

\begin{abstract}
	Кількісно і аналітично досліджено динамічні характеристики розповсюдження інфекції через гетерогенну популяцію з використанням динамічної версії стохастичної спрйнятливо-виявленої-інфекційно-усунутої епідемічної моделі (SEIR), що базується на моделі Ваттса і Строгаца мереж ``малого світу''. Ця модель включає в себе основні реалістичні параметри, що зазвичай характеризують трансмісійні хвороби, такі як сила інфекції, латентність і масові періоди інфекції. Оскільки періоди латентності залишаються меншими, ніж періоди інфікування, частка інфікованих осіб збільшується експоненціально з часом, а в іншому випадку з'являється осциляторна поведінка. Цим можна пояснити періодичність характеристик в часі, що спостерігаються службами запобігання захворювань. Показано також, що періодичні епідеміологічні дослідження переоцінюють або недооцінюють динаміку інфекції, якщо періоди обстеження точно не відповідають характерним часом інфекції. Подано подальше обговорення процесів розповсюдження та релаксації, що беруть участь у цій моделі, та їхнього відношення до характеристичного часу інфекції.
	\keywords мережа ``малого світу'', сприйнятливо-виявлена-інфекційно-усунута епідемічна модель~(SEIR), час зараження, час латентності розповсюдження, епідеміологічні дослідження, дифузійні та релаксаційні процеси
\end{abstract}


\begin{thebibliography}{99}

\bibitem{1} Fraser D.W., Tsai T.R.,  Orenstein W., Parkin W.E., Beecham H.J.,  Sharrar R.G., Harris J., Mallison G.F., Martin~S.M., McDade J.E., Shepard C.C., Brachman P.S., N. Engl. J. Med., 1977, \textbf{297}, 1189, \\ \doi{10.1056/NEJM197712012972201}.

\bibitem{2}  McDade J.E., Shepard C.C., Fraser D.W., Tsai T.R., Redus M.A., Dowdle W.R., N. Engl. J. Med., 1977, \textbf{297}, 1197,  \doi{10.1056/NEJM197712012972202}.

\bibitem{3} Hayes E.B., Plesman J., N. Engl. J. Med., 2003, \textbf{348}, 2424, \doi{10.1056/NEJMra021397}.


\bibitem{3a} Plesman J., Int. J. Microbiol., 2006, \textbf{296}, 17, \doi{10.1016/j.ijmm.2005.11.007}.

\bibitem{4} Robinson M.C., Trans. R. Soc. Trop. Med. Hyg., 1955, \textbf{49}, 28, \doi{10.1016/0035-9203(55)90080-8}.

\bibitem{4a} Saluzzo J.P., Cornet M., Digoutte J.P., Dakar Med., 1983, \textbf{28}, 497.

\bibitem{5}  Ripert C., Le Coeur S., Kanshana S., Jourdain G., Med. Trop., 2003, \textbf{63}, 381.

\bibitem{6}  Garten R.J., Davis C.T., Russell C.A., Shu B., Lindstrom S., Balish A., Sessions W.M., Xu X., Skepner E., Deyde V., \textsl{et al.}, Science, 2009, \textbf{325}, 197, \doi{10.1126/science.1176225}.

\bibitem{7} D'Alembert J. le R., In: Opuscules Mathematiques, Vol.~2, D'Alembert J. le R. (Ed.), David, Paris, 1761, 26.

\bibitem{8}  Dietz K., Heesterbeek J.A.P., Math. Biosc., 2002, \textbf{180}, 1--21, \doi{10.1016/S0025-5564(02)00122-0}.

\bibitem{9}  Kermack W.O., McKendrick A.G., Proc. R. Soc. London, Ser. A, 1927, \textbf{115}, 700, \doi{10.1098/rspa.1927.0118}.

\bibitem{10}  Anderson R.M., May R.M., Infectious Diseas of Humans: Dynamics and Control, Oxford University Press, Oxford, 1992.

\bibitem{11} Keeling M.J., Rohani P., Modeling Infectious Diseases in Humans and Animals, Princeton University Press, Princeton, 2008. 

\bibitem{12} Strogatz S.H., Nature, 2001, \textbf{410}, 268, \doi{10.1038/35065725}.

\bibitem{13} Kleczkowski A., Ole\'{s} K., Gudowska-Nowak E., Gilligan C.A., J. R. Soc., Interface, 2001, \textbf{9}, 158, \\ \doi{10.1098/rsif.2011.0216}.

\bibitem{14} Hartvigsen G., Dresch J.M., Zielinski A.L., Macula A.J., Leary C.C., J. Theor. Biol., 2007,  \textbf{246}, 205, \\ \doi{10.1016/j.jtbi.2006.12.027}.

\bibitem{15}  Luo W., Int. J. Health Geographics, 2016, \textbf{15}, 28, \doi{10.1186/s12942-016-0059-3}.

\bibitem{16} Ma J., van den Driessche P., Willeboordse F.H., J. Theor. Biol., 2013, \textbf{325}, 12, \doi{10.1016/j.jtbi.2013.01.006}.

\bibitem{17} Maharaj S., Kleczkowski A., BMC Public Health, 2012, \textbf{12}, 679, \doi{10.1186/1471-2458-12-679}.

\bibitem{18} Reppas A.I., Spiliotis K., Siettos C.I., Landes Bioscience Virulence, 2012, \textbf{3}, 146, \doi{10.4161/viru.19131}.

\bibitem{19} Castillo-Chavez C., Bichara D., Morin B.R., Proc. Natl. Acad. Sci. U. S. A., 2016, \textbf{113}, 14582, \\ \doi{10.1073/pnas.1604994113}.

\bibitem{20}  Enns E.A., Brandeau M.L., J. Theor. Biol., 2015, \textbf{371}, 154, \doi{10.1016/j.jtbi.2015.02.005}.

\bibitem{21}  Wang H., Li Q., D’Agostino G., Havlin S., Stanley H.E., Van Mieghem P., Phys. Rev. E, 2013, \textbf{88}, 022801, \\ \doi{10.1103/PhysRevE.88.022801}.

\bibitem{22}  Vazquez A., Phys. Rev. E, 2006, \textbf{74}, 056101, \doi{10.1103/PhysRevE.74.056101}.

\bibitem{23}  Zekri N., Clerc J.P., Phys. Rev.~E, 2001, \textbf{64}, 056115, \doi{10.1103/PhysRevE.64.056115}.

\bibitem{24}  Stauffer D., Aharony A., Introduction to Percolation Theory, Taylor \&  Francis, London, 1992.

\bibitem{25} Ochab J.K., G\'{o}ra P.F., Eur. Phys. J. B, 2011, \textbf{81}, 373, \doi{10.1140/epjb/e2011-10975-6}.

\bibitem{26}  McCluskey C.C., Nonlinear Anal. Real World Appl., 2010, \textbf{11}, 55, \doi{10.1016/j.nonrwa.2008.10.014}.

\bibitem{27}  Younsi F.-Z., Bounnekar A., Hamdadou D., Boussaid O., Tsinghua Sci. Technol., 2015, \textbf{20}, 460, \\ \doi{10.1109/TST.2015.7297745}. 

\bibitem{28}  Direction de la Sant\'e et la Population, Oran, Algeria, 2012, private communication.

\bibitem{29}  Newman M.E.J., Watts D.J., Phys. Rev. E, 1999, \textbf{60}, 7332, \doi{10.1103/PhysRevE.60.7332}.

\bibitem{30}  Kretschmar M., Morris M., Math. Biosci., 1996, \textbf{133}, 165, \doi{10.1016/0025-5564(95)00093-3}.

\bibitem{31}  Albert R., Jeong H., Barab\'{a}si A-L., Nature, 1999, \textbf{401}, 130--131, \doi{10.1038/43601}.

\bibitem{32}  Barab\'{a}si A.-L., Albert R., Science, 1999, \textbf{286}, 509--512, \doi{10.1126/science.286.5439.509}.


\bibitem{34} Zekri N., Clerc J.P., C. R. Phys., 2002, \textbf{3}, 741, \doi{10.1016/S1631-0705(02)01367-1}.

\bibitem{33}  Landau L.D., Lifshitz E.M., Statistical Physics, Vol.~5, Elsevier, North Holland, 2013.

\bibitem{35} Communicable disease epidemiological. Central African Republic and Chad 
profile, \\ WHO/HSE/GAR/DCE/2009.2, World Health Organization, 2010. 
\\ URL~\url{https://www.who.int/diseasecontrol_emergencies/toolkits/chad/en/}.

\bibitem{36} Metzler R., Barkai E., Klafter J., Phys. Rev. Lett., 1999, \textbf{82}, 3563, \doi{10.1103/PhysRevLett.82.3563}. 

\bibitem{37} Metzler R., Klafter J., Phys. Rep., 2000, \textbf{339}, 1, \doi{10.1016/S0370-1573(00)00070-3}.

\bibitem{38}  Kremer F., Shonhals A., Broadband Dielectric Spectroscopy, Springer, Heidelberg, 2003.

\bibitem{39}  Heymans N., Bauwens J.-C., Rheol. Acta, 1994, \textbf{33}, 210, \doi{10.1007/BF00437306}.

\end{thebibliography}
\end{document}